\documentclass[amssymb, reprint, showpacs, aip, floatfix, nofootinbib]{revtex4-1}

\usepackage{graphicx}			
\usepackage{dcolumn}			
\usepackage{bm}					
\usepackage[colorlinks=true,linkcolor=blue,citecolor=blue,urlcolor=blue]{hyperref}
\usepackage{float}			
\usepackage{amsmath,amsfonts}	

\usepackage{setspace}		

\usepackage{lineno}   			
\usepackage{ragged2e}
\usepackage{subcaption}
\DeclareCaptionJustification{justified}{\justifying}
\usepackage{multirow}
\usepackage{makecell}

\usepackage{units}

\usepackage[dvipsnames]{xcolor}

\begin{document}


\title{Exploring Quality and Generalizability in Parameterized Neural Audio Effects}

\author{William Mitchell}
\email{williammitchell5818@gmail.com}
\affiliation{Department of Chemistry \& Physics,  Belmont University, Nashville, TN USA}
\author{Scott H. Hawley}
\affiliation{Department of Chemistry \& Physics,  Belmont University, Nashville, TN USA}



\date{\today}

\begin{abstract}
Deep neural networks have shown promise for music audio signal processing applications, often surpassing prior approaches, particularly as end-to-end models in the waveform domain. Yet results to date have tended to be constrained by low sample rates, noise, narrow domains of signal types, and/or lack of parameterized controls (i.e. “knobs”), making their suitability for professional audio engineering workflows still lacking. 
This work expands on prior research published on modeling nonlinear time-dependent signal processing effects associated with music production by means of a deep neural network \cite{signaltrain},  
one which includes the ability to emulate the parameterized settings you would see on an analog piece of equipment, with the goal of eventually producing commercially viable, high quality audio, i.e. 44.1kHz sampling rate at 16-bit resolution.
The results in this paper highlight progress in modeling these effects through architecture and optimization changes, towards increasing computational efficiency, lowering signal-to-noise ratio, and extending to a larger variety of nonlinear audio effects. Toward these ends, the strategies employed involved a three-pronged approach: model speed, model accuracy, and model generalizability. 
Most of the presented methods provide marginal or no increase in output accuracy over the original model, with the exception of dataset manipulation. We found that limiting the audio content of the dataset, for example using datasets of just a single instrument, provided a significant improvement in model accuracy over models trained on more general datasets.  
\end{abstract}

\maketitle

\section{\label{sec:1} Introduction}
Computer technology and modeling has exploded over the last decade with the improvement in computer processing power and techniques. With these advances in speed and power, along with the advent of using graphics processing units (GPU) for batch computation, machine learning processes and neural network architectures have begun to tackle increasingly rigorous tasks and problems in almost every major research field from medicine \cite{medicine}, to justice systems \cite{justice}, to audio processing \cite{open_source_music,marco_equalization,SING}. However, neural network and deep learning processes are still a relatively new research field, and better training techniques, architecture designs, loss functions and learning processes are introduced to the academic community regularly. These features can then be further developed in new fields to determine their optimal purpose. Modeling audio effects is an especially interesting field of research because of both its complexity and its usefulness across multiple disciplines.

Being able to accurately and efficiently model audio effects provides for numerable advantages including portability, repeatability, and flexibility. Modeling analog audio effects in the digital domain allows those effects to be increasingly portable, they require no physical space or weight and can be uploaded onto numerous devices in numerous locations. Digital effects models also provide greater repeatability over analog effects, which may require calibration or experience degradation over time. Finally, digital effect models can be more flexible, with a greater range of input formats and an easier opportunity for modification. Similar state-of-the-art research in this field include a variety of model types and practices including generative adversarial networks \cite{gan1,gan2,gan3}, and multiple types of model autoencoders. For further discussion of relevant applications  see ``Autoencoders for Music Sound Modeling''\cite{autoencoders}.

The research in this paper presents the progress made on already published results\cite{signaltrain} on modeling nonlinear time-dependent signal processing effects associated with music production by means of a deep neural network in the waveform domain. This research has a specific focus on improving the speed, accuracy, and generalizability of our previously published SignalTrain model \cite{signaltrain} on nonlinear effects. Linear systems possess two mathematical properties: homogeneity and additivity. If a given system or effect doesn’t have one, or both, properties the effect is considered nonlinear. Examples of these effects are things like compression and distortion. A majority of the past research in this field utilizes spectrograms, time and frequency graphs of audio using the Fourier transform, of input and output audio to train the network. The waveform domain was considered too computationally expensive, but it's usefulness has been explored more in recent years \cite{waveform_domain1,waveform_domain2}. This work focuses on the waveform domain because of the ability to retain both the frequency and phase information of the audio, something that spectrograms cannot do, in order to improve both the accuracy of training and the quality of output audio. This work also includes the application of trainable virtual “knobs” into the neural network architecture, which act as the virtual version of analog knobs that can control various settings on analog effects units. Much of the following results are from compressor effects specifically, mostly because it was found to be the hardest effect to model, but we also show that the model is trainable on a wide array of effects, both digital and analog. This area of research is significant because it presents a harder computational task than both categorizing and recognizing audio content, by being able to reproduce a specific effect that will produce a desired change in the audio content.

Audio effect modeling is also an increasingly lucrative research area, with many commercially viable applications. Current commercial products such as the Kemper Profiler \cite{kemper} and Fractal Fx-II \cite{fractal}, are leading the modern surge of commercial audio modeling. The ability for the general populace to have access to more and more powerful systems has created an opportunity for research in this field to become immediately usable and practical in a commercial sense. Advanced knowledge in these tasks will not only help further the understanding and handling of audio waveforms and files but will help further the understanding of how to better optimize, train, and structure neural network models for similar tasks. Learning and experiencing how neural networks extract features and detect patterns in waveforms across these audio processes will inform both the audio technology and machine learning fields. Our research goal is to utilize the advancing power and adaptability of neural networks to effectively model commonly used audio effects (i.e. compression, echo, reverb), by training a neural network architecture on pre- and post-effect audio waveforms, along with a focus on high quality output audio. All the following presented research can be applied to any quality of input audio, but this work focuses on high-quality CD level audio, i.e. 44.1kHz sampling rate with a 16-bit depth. Following is presented an overview of the SignalTrain architecture, along with the researched methods for improving the three areas of interest: speed, accuracy, and generalizability.

\section{\label{sec:methods} Methods}

\begin{figure}[]
\includegraphics[width=0.9\columnwidth]{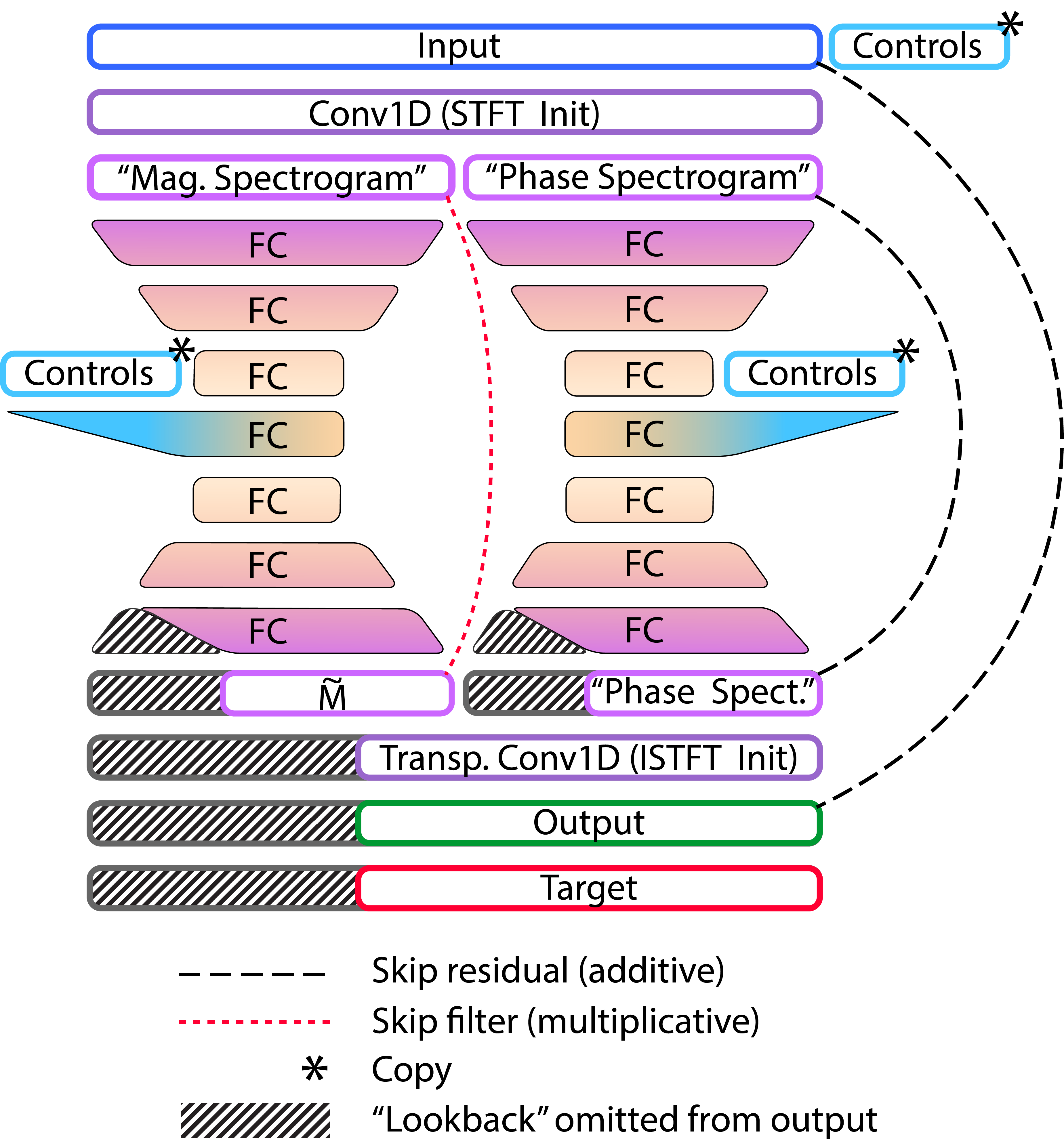}
\caption{\label{fig:1}{
Overview of the original architecture of the SignalTrain model \cite{signaltrain}.
}}
\label{fig:model}
\end{figure}

SignalTrain implements a deep neural network, with architecture inspired from U-Net \cite{UNET} and TFNet \cite{TFNET}. Like U-Net, SignalTrain utilizes an “hourglass” encoder-decoder architecture with skip connections spanning across the middle. Like TF-Net, SignalTrain also works in both the time and spectral domains explicitly, as outlined in the model overview presented in Figure \ref{fig:model}. Most of the original model architecture remains constant throughout the results presented in this paper, with very minor experimental changes made that are detailed below. The code used is
the original SignalTrain source\footnote{Source code at \url{http://github.com/drscotthawley/signaltrain}},
subject to modifications that follow.
The front-end module contains two 1-D convolution operators that produce a single sub-space that provides magnitude and phase features. These features are processed individually by two following deep neural networks which comprise the autoencoder module. These networks contain 7 fully connected, feed-forward neural networks, who are also conditioned by the control variables, or “knobs”, of the audio effect module. The SignalTrain model learns a mapping of the un-processed to the processed audio, by the audio effect to be profiled, and is conditioned on the vector of the effect’s controls. For a more complete description of the model see \cite{signaltrain}.

Datasets in the following experiments are made up of appropriate audio data taken from either pre-recorded samples through the creative commons license or are personally manufactured recorded samples using popular analog audio effects processors like the Universal Audio LA2A \cite{LA2A}. Digital effects courtesy of Dr. Eric Tarr’s HackAudio digital effects, implemented on music only datasets of amateur songs via the creative commons license, along with established datasets like the IDMT-SMT-Audio-Effects dataset \cite{IDMT_SMT} are used to learn both digital-only and analog effects. This specific dataset was also used in \cite{marco_blackbox_modeling}, which has been taken advantage of to provide a direct comparison. This dataset provided audio samples from a Leslie Cabinet speaker and a Universal Audio 6176 Vintage channel strip. Modeling some of the most successful analog equipment will help to better understand both why they might have been successful, and how different circuitry designs can affect processing in unexpected ways. Similar combinations of the popular FMA dataset \cite{FMA} and the NSynth dataset \cite{neural_audio_synthesis} were used in this experiment and any audio used not already contained in a previously constructed dataset can be found on Zenodo under the "SignalTrain Concatenated Dataset" \cite{signaltrain_concatenated}. It is expected that the power of deep neural networks using raw audio waveforms will provide the ability to model a myriad of nonlinear audio processes, not simply one facet such as compression or echo generation. After training on a dataset, new audio can be manufactured from the trained model checkpoint and be compared to the original audio.

Many of the parameters of the model and datasets were subject to change and are detailed in the following sections. Unless otherwise specified, a log-cosh loss function \cite{logcosh} was used across all trainings and weight initialization was randomized except for input/output rates, which were initialized using a discrete Fourier Transform \cite{fourier_transform}. In use was a PyTorch-based model utilizing raw audio waveforms as input and output data on a single NVIDIA RTX2080Ti GPU with mixed precision training. The remaining parameters were altered in order to further our goals in a three-pronged approach: 1) improving computation speed, 2) improving overall model accuracy or, 3) lowering the signal to noise ratio of the output audio to achieve higher quality outputs. Altering these parameters gives a deeper insight into how the network is functioning and learning, which proved valuable in optimizing its performance.

\section[title]{Experimentation/Results\protect\footnotemark[2]\label{sec:experiment}}
\footnotetext[2]{Audio examples of results are available at\hfill\break 
\url{https://tinyurl.com/signaltrain-exploring}}

\subsection{Speed}
\label{sec:speed}
The first area of interest of the three-pronged approach to improving the model was increasing its efficiency, or decreasing the time it required to train on data while still achieving comparable accuracy. Methods that improve accuracy often increase training time, and those that decrease training time often bring a decrease in accuracy too. Striking the right balance between speed and accuracy is often dependent on the situation and for each following method we discuss that balance. A model, identical to that described in \cite{signaltrain}, was used to create baseline results for comparison. This model was trained on 200,000 datapoints of 5-second 16-bit 44.1kHz audio, for 1,000 epochs using a compressor effect with 4 variable knobs on a NVIDIA RTX 2080 GPU. Typical runs for the baseline averaged between 12-13 hours, and a single training example that lasted 12.32 hours and achieved a training loss of 
4.899e-06 is used for direct comparison results.

\begin{figure}[]
\includegraphics[width=0.9\columnwidth]{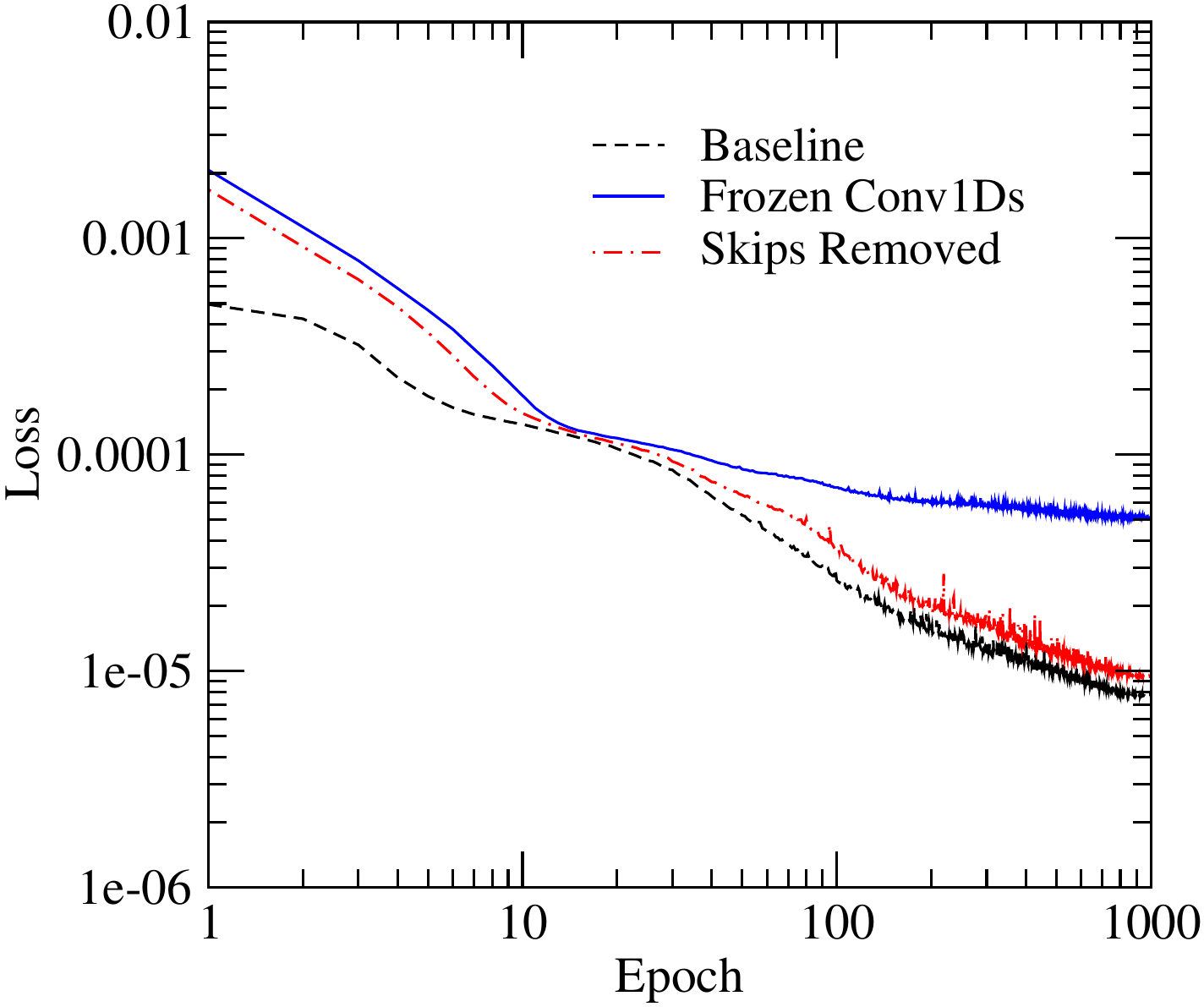}
\caption{\label{fig:2}{
Effects of changes to model architecture. Log-Log graph comparison of the validation set loss values for the baseline, frozen transform layers (Conv1D layers in Figure 1), and removed final skip connection.  Note that the `hump' shapes in this and later plots results from the system's use of a ``1cycle'' learning rate schedule\cite{signaltrain}.}}
\end{figure}

The first attempt at improving the speed of the model was freezing transform layers. Fourier transform layers are used to go to and from spectrograms twice in the model, from input to output audio. These layers are initialized with Fourier weights, but become learnable and are updated throughout the training procedure. Without having to do the extra computations to update these weights, the model would train faster. Shown in Table \ref{Table:1}, freezing the Fourier transform weights decreased model training time significantly, by about 11.56 percent, but accuracy worsened by a factor of 8.72, or the loss value was 8.72 times worse. A log-log graph with the validation loss values are displayed in Figure \ref{fig:2}. While the model did become faster, such a large increase in error was considered too great and continuing to freeze layers was not performed.

\vspace{10pt}

\begin{table}[]
\resizebox{\columnwidth}{!}{%
    \begin{tabular}{l|l|l}
    \thead{Description} & \thead{Training\\ Time (Hours)} & \thead{Validation\\Loss} \\
        \hline \hline
      Baseline & 12.81 & 7.666e-6 \\
       \hline
     Frozen Layers & 10.90 & 5.128e-5 \\
      \hline
     No Skip Connection & 12.81 & 9.000e-6\\
     \hline
    \end{tabular}
}
    \caption{\label{Table:1}Comparison between a baseline SignalTrain model, one with frozen Fourier Transform layers, and one with removed final skip connections. All trained on the same dataset.}
\end{table}

The second attempt at increasing model efficiency was by analyzing the effect of skip connections across the model. Skip connections are widely considered to be purely beneficial to neural network training \cite{skip_connections1,skip_connections2}, but their applicability across different deep learning problems is variable. Also shown in Table \ref{Table:1} are the results from removing the final skip connection, between the final output and initial input. As shown, including these skip connections adds essentially zero time to the training process but improves the model’s training loss value by 24.4 percent, clearly a model design that is purely beneficial in this application. Figure \ref{fig:2} also presents the validation loss results of the removed skip connection trained model.


\subsection{Accuracy}
\label{sec:accuracy}

The second area of interest for improving the model was to improve how accurate the model could become, or how well the model could learn to mimic the desired effect. As mentioned earlier accuracy and speed often compete when implementing model changes and becoming more accurate may include increasing the computational effort, which will increase model training time. The first result listed in Table \ref{Table:2} presents results of training the same baseline model used for 10 times more epochs. 

\begin{table}[h]
\resizebox{\columnwidth}{!}{%
    \begin{tabular}{l|l|l}
    \thead{Description} & \thead{Training\\ Time (Hours)} & \thead{Validation\\Loss} \\
        \hline \hline
      Baseline (1,000 epochs) & 12.81 & 7.666e-6 \\
       \hline
     10,000 Epochs & 131.6 & 3.540e-6 \\
      \hline
     Vocal Only Dataset & 12.54 & 2.838e-6\\
     \hline
     Vocal Dataset 16kHz & 11.57 & 5.425e-6\\
     \hline
    \end{tabular}
}
    \caption{\label{Table:2} Comparison of the baseline SignalTrain model, a model trained on 10,000 epochs, and two models trained on a dataset containing only vocal audio.}
\end{table}

\begin{figure}[]
\includegraphics[width=0.9\columnwidth]{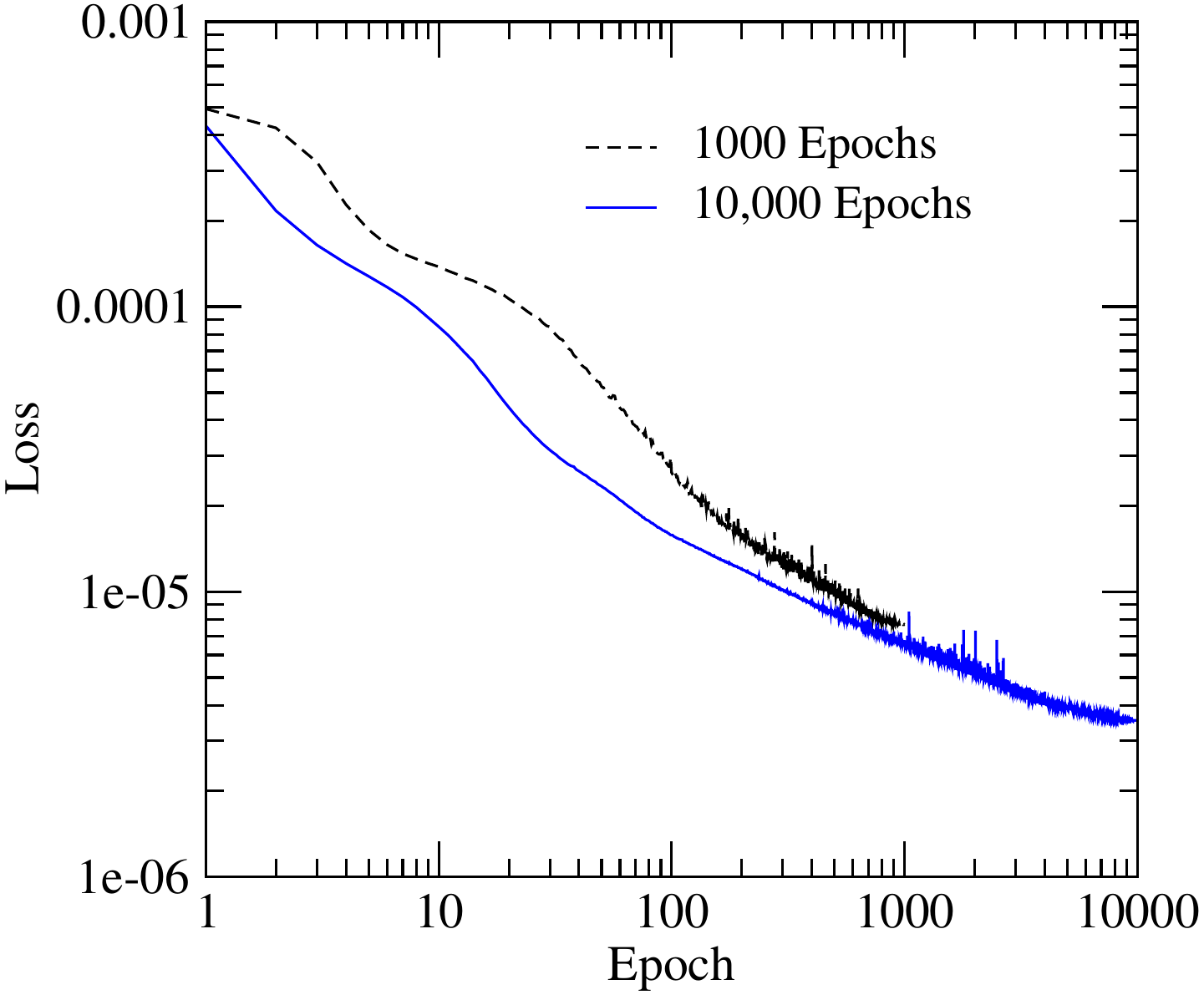}
\caption{\label{fig:4}{
Effect of longer training. Log-Log graph comparison of the validation set loss values of the baseline model run for 1,000 vs. 10,000 epochs. Although the loss is lower and audio quality better for 10,000 vs 1000 epochs, the increase in quality seems disproportionately small compared to the factor of 10 increase time.}}
\end{figure}

This model again used 200,000 16-bit, 44.1kHz 5-second audio datapoints on a compressor effect with 4 variable knobs, but this time trained for 10,000 epochs instead of 1,000. This was performed primarily to test the limits of the model, and how low the loss value could become given more time to learn. As shown, the loss value of this extended training reached 3.826e-06, just marginally better than the baseline model, even with 131.6 hours of training. This indicates training for significantly longer than 1,000 epochs will produce negligibly better results. While one could train the model for significantly longer (e.g., an entire month), it is not obvious that this alone would provide the increase in quality suitable for pro audio workflows -- and this was for a 'small' input window size of only 4096 samples; larger windows would require significantly more expenditures of computational resources, beyond the scope of this academic study. Figure \ref{fig:4} shows the log-log graph of the validation loss values between these two trainings.

\begin{figure}[]
\includegraphics[width=0.9\columnwidth]{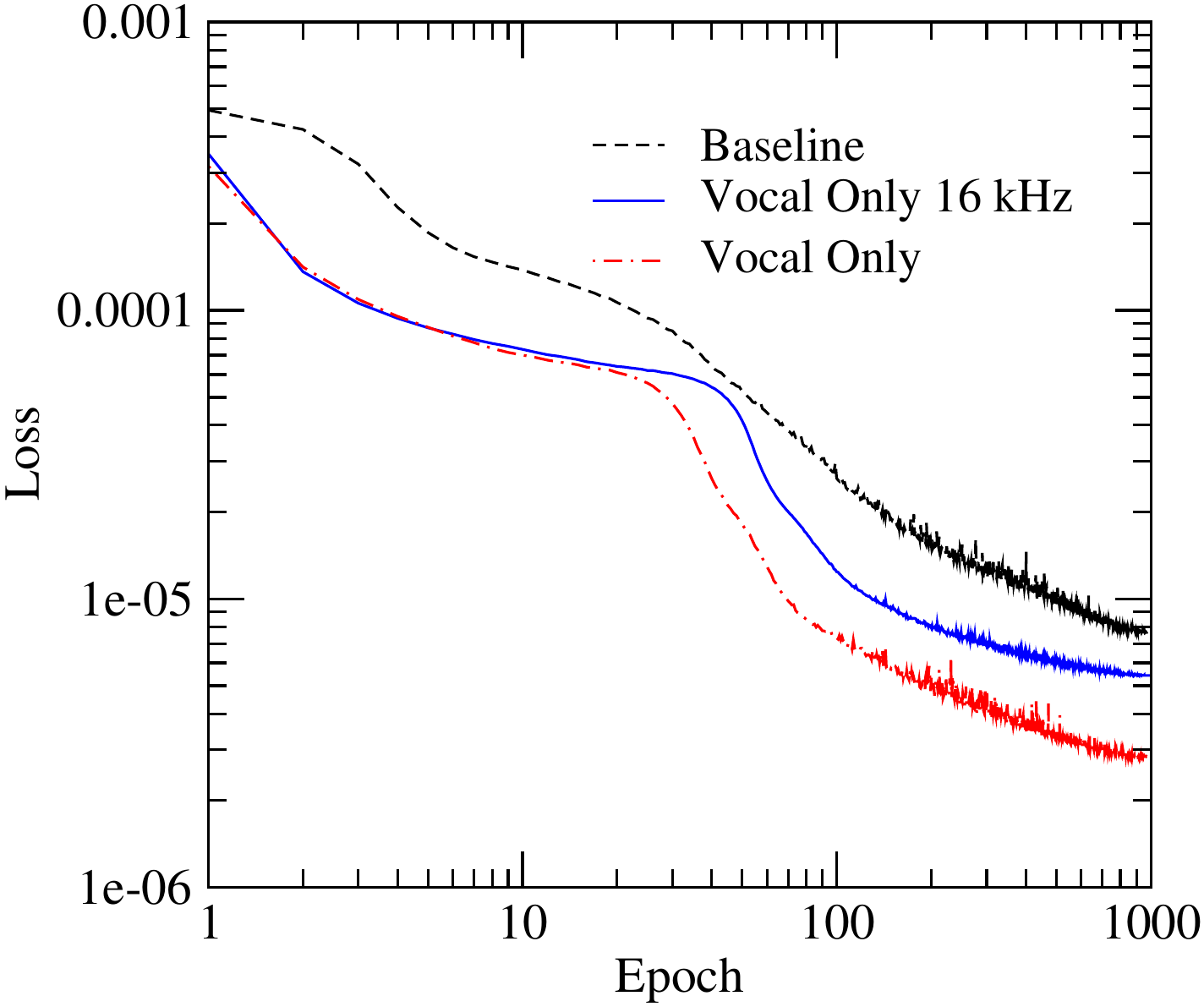}
\caption{\label{fig:5}{
Effects of restricting the dataset.  Log-Log graph comparison of the validation set loss values of the baseline dataset (consisting of a variety of sounds), a dataset of vocals only, and a vocal-only dataset with sampled to 16kHz.}}
\vspace{-10pt}
\end{figure}

Another attempt at improving the accuracy of the model was to restrict the variability of the audio used to construct datasets. Previous work utilized datasets of randomly generated audio, and the baseline described in this paper utilized amateur recordings of music of various types, with no restrictions on genre, instrumentation, or amplitude. Shown in Table \ref{Table:2}, restricting datasets to single instruments, in this case only vocals, proved beneficial in lowering the loss value. The vocal only dataset used was constructed using amateur recordings of acapella vocals, in multiple languages via the creative commons license. These same recordings were then down sampled to 16kHz to create an identical dataset, just with 16kHz sampled audio. Figure \ref{fig:5} displays a log-log graph comparing the baseline, vocal only, and vocal only 16kHz sampled validation loss results. The vocal-only run achieved a training loss value 46.95 percent better than the baseline, and the 16kHz sampled audio achieved a training loss value 11.1 percent worse than the baseline, although the validation loss was better than baseline, and it decreased training time by just over a half hour. 

\begin{figure}[]
\includegraphics[width=1\columnwidth]{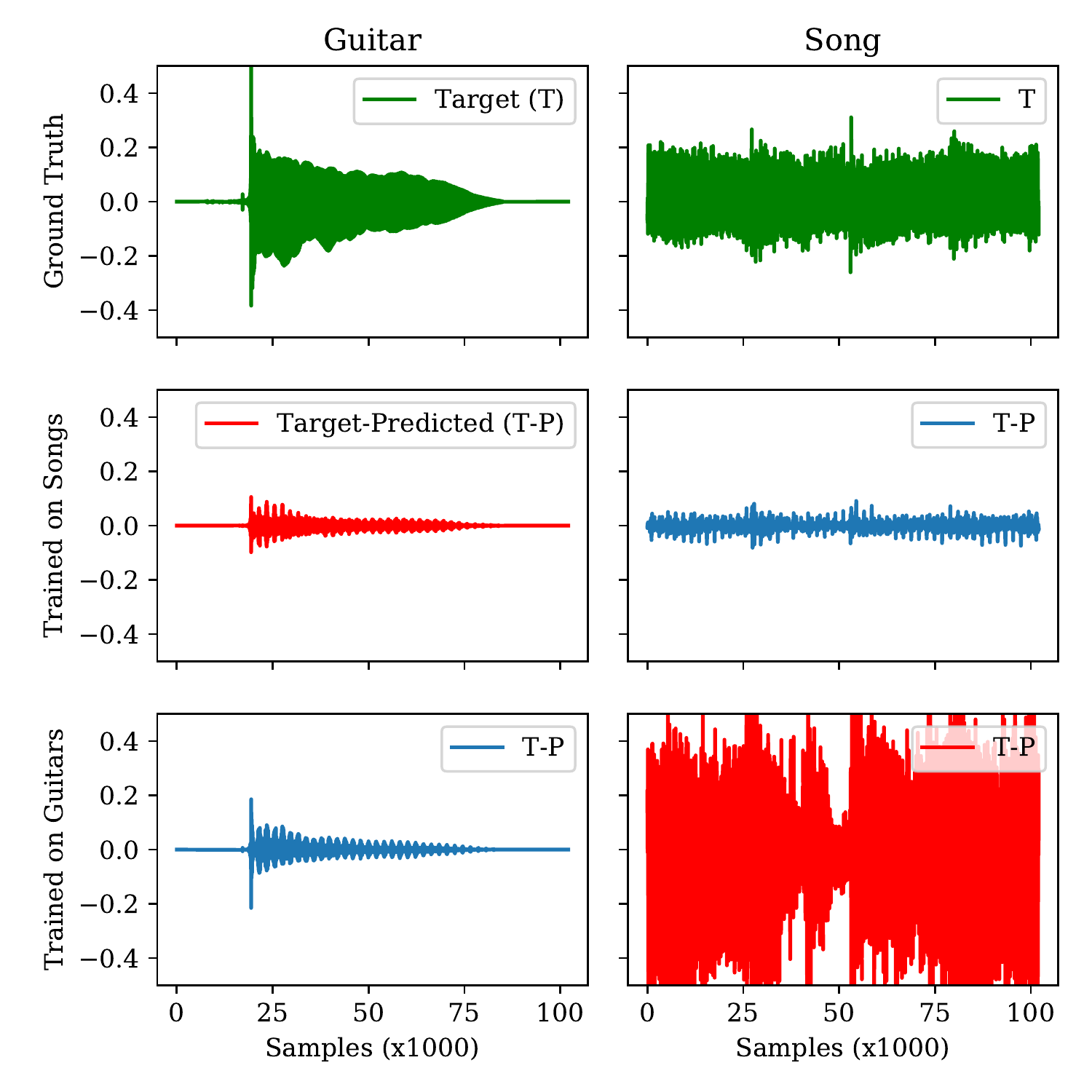}
\caption{\label{fig:waveform_comparison}{
Waveforms showing guitar note (left column) and a portion of a full-band song (right column) for the comp4c compressor effect.  Top row: Target audio.  Middle row: Difference between target and predictions made by a model trained on songs.  Bottom row: Difference between target and predictions of a model trained on guitar notes. Of the two lower rows, those on the main diagonal (also shown in red) are models whose training dataset was not of the same type as the prediction, whereas the others (shown in blue) reflect training datasets similar to the prediction.   Note that the model trained on guitar notes fails to generalize to the full song (bottom right). Despite the appearance of plots in the left column, the model trained on guitars (bottom, blue) actually predicts the guitar note with lower log-cosh error and less audible noise than the model trained on songs (middle, red). Audio examples 
are available at \url{https://tinyurl.com/signaltrain-exploring}.
}}
\vspace{-10pt}
\end{figure}

Improving the accuracy of model training is important because it should directly correlate to improved audio output quality, although this is not necessarily always the case. Qualitatively, varying aspects of the training dataset will improve output audio quality with relatively no change in loss value. Current research indicates that increased amount of silence in the dataset will correspond to more noise in output audio without worsening the loss value. Similarly, datasets with more restricted contents do not generalize as well to other types of audio compared to those datasets that contain more diverse contents. For example, models trained on single guitar notes produce better sounding outputs when given only guitar notes than when given audio samples of full songs.  Figure \ref{fig:waveform_comparison} shows waveforms demonstrating this phenomenon. In the lower right plot one sees that the predicted output contains severe distortion covering essentially the entire waveform, indicating that models trained only on guitar sounds will not generalize well to predicting full-band songs. 

In addition to the approaches already mentioned, we also investigated alternative loss functions, such the log-cosh of the difference in output spectrogram values (as opposed to the time-domain values), and a log-SNR loss\cite{mim20_uirl_eusipco}, but were unable to observe improvements to the accuracy compared to the baseline.

\subsection{Generalization: Extending to Various Audio Effects}
\label{sec:generalization}

Up to this point in the paper, we have only considered improvements for modeling a compressor effect. The final area of interest was exploring the ability of SignalTrain to model various different audio effects beyond compressors. Previous work focused mainly on digital compressors because of their hard-to-model nature, although nothing required the modeled effect had to be a compressor, much less a digital compressor. The first progress in expanding the model’s generalizability was to include effects that differed significantly from compressors, and to extend further into the analog domain. To achieve this the freely available AudioMDPI dataset \cite{marco_blackbox_modeling}, was used. This dataset contains single guitar or bass notes, lasting 2-seconds in length from the IDMT-SFT-Audio-Effects dataset \cite{IDMT_SMT}, as their dry/input samples. These dry samples were then processed through a Leslie cabinet speaker \cite{leslie} horn and woofer separately, and a Universal Audio 6176 Vintage channel strip.

Table \ref{Table:3} presents the training results on these datasets over 1,000 epochs on the chorus and tremolo effects. It should be noted that the increased training times were due to an increase in input audio buffer size, which was necessary due to the nature of the effects. Compressors, as an effect, do not introduce any oscillatory characteristics into the audio, however Leslie cabinets have rotating horns that oscillate with some set frequency. 
One may expect the Chorus effects to be more computationally intensive for our model to reproduce than the Tremelo effects, as the former have rotation periods approximately 8 times longer than the former, requiring proportionately larger input buffer sizes to accommodate.
Initial experimentation demonstrated that input buffer sizes that did not contain a complete cycle of the horn could not model the effect with any sort of accuracy. Increasing the input buffer size from 8192 to 98304 samples proved sufficient with these datasets. 

\begin{table}[ht]
\resizebox{\columnwidth}{!}{%
    \begin{tabular}{l|l|l}
    \thead{Description} & \thead{Training\\ Time (Hours)} & \thead{Validation\\Loss} \\
        \hline \hline
      Baseline (1,000 epochs) & 12.81 & 7.666e-6 \\
       \hline
     Leslie Horn Chorus & 48.15 & 6.817e-5 \\
      \hline
     Leslie Woofer Chorus & 51.01 & 4.893e-5\\
     \hline
     Leslie Horn Tremolo & 48.33 & 2.104e-4\\
     \hline
     Leslie Woofer Tremolo & 51.35 & 1.225e-4\\
      \hline
    \end{tabular}
}
    \caption{\label{Table:3}Comparison between our baseline model and models trained on various analog effects from a Leslie Cabinet. Increased training times are due to the increased input chunk size to account for the oscillatory nature of these effects.}
\end{table}

It should be noted that the validation loss results presented in Table \ref{Table:3} present the final loss value and not necessarily the lowest value achieved during training. It was found that the model experienced significant over-fitting during the training of the Leslie effects, most likely due to the relatively small size of the dataset. The AudioMDPI dataset contains approximately 251MB of data per effect, and the datasets used to model the digital compressor effects discussed previously average near 30GB of data. This indicates the SignalTrain model requires much larger datasets than AudioMDPI to achieve qualitatively good results, a characteristic that can be further defined in future research.

\section{\label{sec:conclusion} Conclusion and Future Work}
This paper presents the improvement and optimization work done to make the SignalTrain model more efficient, more accurate, and for the purposes of better audio quality. First, it explained attempted methods for trying to improve the efficiency of the model, which would decrease overall training time. It was found that freezing Fourier transform layers would improve speed, but not to the extent that it would justify the great loss in model accuracy. It was also shown that the skip connections in the original model do not increase training time and do significantly help the model achieve lower loss values. Next it presented the results of attempting to increase the model’s accuracy through allowing the model to train for significantly longer, and by restricting the variability of the datasets used for training. It was shown that for runs substantially over 1,000 epochs, the model does improve its loss value, but only marginally and no noticeable improvement in output audio quality was observed. It was also shown that restricting datasets to single instruments improved the model training significantly, without increasing training time. However, models trained on single instruments failed to generalize to other types of audio and were only effective when given instruments that it was trained on. Finally, the results at expanding the generalizability of the SignalTrain model were shown. Both analog (Leslie Cabinet) and digital (HackAudio compressors) effects were modeled effectively. These effects also varied significantly in how they alter audio, proving the SignalTrain model can learn more than just compressors. It was also indicated that differing effects require specialized parameters to produce the best results.

There are many potential avenues for future work using the SignalTrain model and in the signal-processing effects modeling field. Future research can be the continued optimization of the SignalTrain model, including specific effect optimization, and further integration of audio plugins within the model. Belmont University, with a large supply of trained listeners, provides an excellent opportunity and space for qualitative listening tests.  Overall, there are many paths and opportunities for the audio machine learning field going forward.

\begin{acknowledgments}
\vspace{-10pt}
William Mitchell wishes to thank Scott H. Hawley for his help and mentorship, and Dr. Eric Tarr, Benjamin Coulburn and Stylianos Ioannis Mimilakis for their prior contributions to SignalTrain.
\end{acknowledgments}

\bibliography{signaltrain2}

\end{document}